\documentclass[twocolumn,showpacs,amsmath,amssymb,floatfix]{revtex4}

\usepackage{graphicx}
\usepackage{dcolumn}
\usepackage{bm}

\begin{document}
\vskip 2cm
\title{
Crystal asymmetry induces single-atom chain formation in gold nanowires
}
\author{Francesca Tavazza$^1$, Anwar Hasmy$^{2,3}$, Lyle E. Levine$^1$, 
Anne M. Chaka$^1$, Luis Rinc\'on$^{2,4}$, M. M\'arquez$^{2,5}$ and Carlos
Gonz\'alez$^2$}
\affiliation{$^1$National Institute of Standards and Technology, Gaithersburg, 
MD 20899, USA}
\affiliation{$^2$ NIST Center for Theoretical and Computational Nanosciences, Gaithersburg,
MD 20899, USA}
\affiliation{$^3$INEST Group Postgraduate Program, Philip Morris USA, Richmond, 
VA 23234, USA}
\affiliation{$^4$Departamento de Qu\'\i mica, Universidad de los Andes, 
M\'erida-5101, Venezuela}
\affiliation{$^5$Research Center, Philip Morris USA, Richmond, 
VA 23234, USA}
\medskip 
\date{\today}
\begin{abstract}
We performed density functional theory and tight-binding molecular 
dynamics calculations to investigate the formation of suspended linear 
atom chains when stretching gold nanowires along the [110] crystal 
orientation. We determined that chain formation can occur only when the 
crystal symmetry is broken in the early stages of the elongation process. 
Such crystallographic asymmetry can be induced by stretching the wire along 
a slightly off-axis tensile direction or by introducing thermal fluctuations.
Our observation of the off-axis formation of these chains agrees with
experimental findings.
\end{abstract} 
\maketitle

\bigskip
\newpage
Nanowires (NWs) exhibit interesting properties that may be 
exploited to generate novel electronic devices\cite{Nanowires}. 
One of the most exciting properties is the observed high stability of 
suspended single-atom chains (SACs) that form during the stretching 
of gold NWs\cite{Ohnishi,Yanson,Ugarte1,Ugarte4,Cadenas3}. The formation
of SACs also suggests that gold NWs may be useful as 
an intrinsic force standard.
However, little is known about the geometrical conditions required for such 
single-atom chain formation. In particular, conflicting experimental results 
have been reported for NWs elongated along the [110] high-symmetry 
axis\cite{Ugarte1,Ugarte4,Cadenas3}.
From the theoretical point of view, density functional theory (DFT)
calculations have failed to describe the formation of 
SACs\cite{SolerTosatti,SanchezPortal}. Due to the high computational cost, 
none of these numerical efforts described the dynamics of the chain formation 
under strain. Other studies have considered molecular dynamics 
simulations of gold SACs based on classical many-body 
empirical potentials\cite{Sorensen,Bahn,AgraitMD,Ugarte2,Ugarte3,HSPark}. 
Although they provided useful insights into the SAC formation process, classical 
approximations are unreliable when quantum effects become relevant, and predictions
of [110] SAC formation depends upon the specific classical potential 
used\cite{Sorensen,Bahn,Ugarte2}.  Da Silva {\it et al.}\cite{Fazzio1,Fazzio2} combined a 
description of the electronic structure and the dynamics of the breaking process 
by using tight-binding molecular dynamics (TBMD) simulations. The probability of 
linear single-atom chain formation, as well the number of atoms within the 
chains, match the experimental observations\cite{Fazzio2}. 
Their work, however, only focused on the dynamics of chain formation along the 
[111] crystallographic orientation. 

In this letter we address the ``[110] controversy,'' by performing
semistatic quantum relaxation calculations using DFT for a first principles description
of the underlying energy landscape and the low-energy atomic configurations, and molecular dynamics
simulations within the NRL-Tight Binding approximation\cite{NRL-TB} to describe the dynamics
effects within the system during deformation.  
To investigate the 
breaking mechanism of [110]-oriented NWs stretched along a [110] tensile axis, 
we considered both the ideal case, where the tensile axis is perfectly aligned 
along the [110] crystallographic direction, and a more realistic one, where 
the crystal symmetry is broken either by appling the strain slightly off-axis or by 
introducing thermal fluctuations. 


The DFT calculations were performed using DMol$^3$ \cite{Dis,D901,D902}. In 
DMol$^3$, the physical 
wave function is expanded in a numerical basis set, and fast convergent 
three-dimensional integration is used to calculate the matrix elements occurring in 
the Ritz variational method. We used a real-space cutoff of 0.4 nm and a double-zeta, 
atom-centered basis set (dnd).
We utilized a generalized gradient approximation (GGA)
approach (Perdew-Burke-Ernzerhof\cite{PBE}), and a hardness conserving semilocal 
pseudopotential (dspp\cite{pseudo}, only electrons with n=5 and n=6 were handled 
explicitly). The geometry optimization was performed using a conjugate gradient approach  
based on a delocalized internal coordinate scheme\cite{geom_optim1,geom_optim2}.
In our DFT calculations we considered NWs with an axis parallel to a [110] crystallographic 
orientation and, initially, atoms in their bulk positions. Three combinations of tensile 
axes were used: [110],  [99~101~2] (about $1^{\circ}$ off the [110]), 
and [110] followed by [99~101~2]. Our simulation cell contained 115 atoms and, to eliminate 
self-interactions between the ends of the chain, a cluster configuration was used. 
Initially, the central part of the NW contained 9 alternating 
planes with 4 or 5 atoms each (see Fig. 1). 
Two atomic layers at the top and bottom served as grips and both single and double sided 
stretching modes were used. The grip atoms were incrementally moved along 
the tensile (z) axis by $\Delta$z = 0.029 nm for the [110] case and along
$\Delta$x=0.00029 nm, $\Delta$y=0.00058 nm and $\Delta$z= 0.029 nm 
for the [99~101~2] case. After each tensile increment, the grip atoms 
were kept fixed while all of the other atoms were allowed to relax into new 
configurations. Usually, the system was considered converged when the change in total 
energy per atom was less than 5 $\times$ 10$^{-6}$ eV and changes in the gradient of 
the free atomic positions were less than 3 $\times$ 10$^{-4}$ eV/$\AA$. 
This methodology has been extensively used in recent years when studying nanowire 
deformation\cite{method}. Different effective strain rates were simulated by 
slightly changing the convergence criteria on the gradient of the atomic positions, 
similar to the procedure in Picaud {\it et al}.\cite{method1}.
\begin{figure}
\centering
\includegraphics[width=7.0cm]{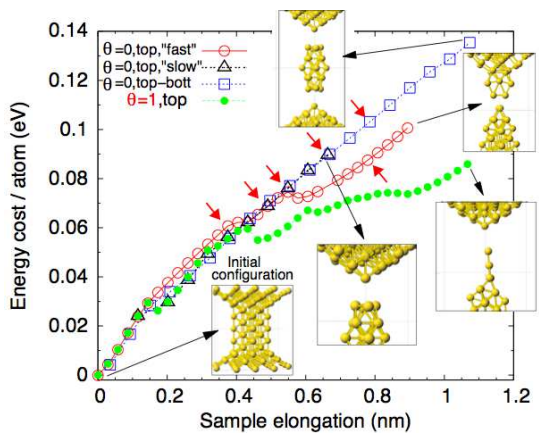}
\caption{\label{fig:energies} Energy cost per atom versus sample elongation 
for selected simulations. 
Deformation along the [110] and [99~101~2] tensile axes is labeled ``$\theta=0$'' 
and ``$\theta=1$'', respectively. The notation``top'' 
indicates that the NW was stretched from one end only, while ``top-bott'' indicates 
symmetrical stretching. Examples of final configurations are 
displayed. The small arrows indicate [110] tensile axis configurations 
later used as starting points for
[99~101~2] tensile axis simulations.}
\end{figure}

The DFT results indicate that initially perfect NWs always break abruptly 
(i.e. simultaneously breaking more than one bond) when stretched along the 
[110] axis, in agreement with the observations of Rodrigues {\it et al}.\cite{Ugarte4}. 
This behavior is caused by the high symmetry of the system where all preferred slip
directions are identical and the system cannot choose one over the
other. Abrupt breaking is found for all effective strain rates and stretching 
modes, even though the energy path and atomic configurations depend on 
those quantities, indicative of a relatively flat yet complex energy surface. 
As a general rule, we find that there is less atomic rearrangement at lower 
effective strain rates. Also, abrupt breaking has a higher 
energy cost than SAC formation. These results are illustrated in Fig. \ref{fig:energies}, where 
the energy cost per atom versus elongation, and the corresponding final structures, 
are displayed for our most representative simulations. Calculations
performed using the [110] tensile axis are indicated by $\theta$=0. Initially,
the NW deforms elastically, and the energy path is identical in all cases. In the 
plastic regime, comparing only the simulations with single-sided stretching
(open circles and triangles), we see that the larger atomic rearrangements at higher 
effective strain rates (open circles) result in larger final elongations ($\approx$ 
0.9 nm compared to $\approx$ 0.7 nm) and higher energies.  
The final force necessary to break a NW pulled along the [110] crystallographic direction 
is very consistent and independent of the strain rate and stretching mechanism:
the average of all our simulations is -2.17 nN$\pm$0.03 nN. 
\begin{figure} 
\centering
\includegraphics[width=6cm]{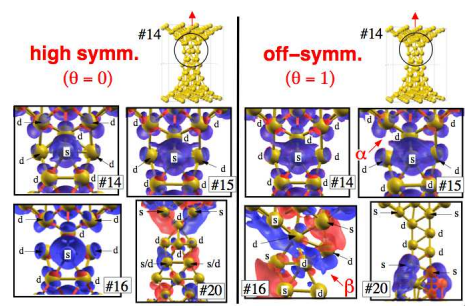}
\caption{\label{fig:evol} Atomic arrangements and HOMO isosurfaces
for crucial steps (14, 15, 16, 20) in the evolution of the NW for tensile
axes [110] ($\theta$=0) and [99~101~2] ($\theta$=1). For each atom, the 
prevailing orbital character is specified (positive and negative lobes are 
colored blue and red, respectivly).  }
\end{figure}
Figure \ref{fig:energies} also shows results from the slightly off-symmetry [99~101~2]
tensile axis (solid circles, $\theta$=1). Here, the NW evolves
basically identically to the high symmetry 
case for elongations up to 0.4 nm. For larger strains, significantly lower 
energy configurations are found, resulting in the formation of a SAC, which 
has also been observed experimentally\cite{Cadenas3}. 

Details of the atomic evolution of the NW and a direct comparison with the high symmetry 
case are displayed in Figure \ref{fig:evol}. Here, the off-axis geometry produces classic 
positive and negative stress concentration points at the base of the NW.
Up to the 14th elongation step, the asymmetric stresses are too small to produce any 
significant differences between the two cases. In the 15th step, the stresses at the 
base of the NW produce an asymmetric bond failure (see arrow $\alpha$) that drastically 
changes the stress distribution and produces further asymmetric bond breaking events 
in subsequent steps (see arrow $\beta$, step 16). The asymmetric deformation of the chain also 
significantly affects the electronic structure of the NW. While the highest occupied 
molecular orbitals (HOMO) are almost identical in the two cases up to the 13th step, 
they start differing in the 14th at the incipient failure point labeled
$\alpha$ in the 15th step.  In the 15th step, the large {\it s} orbital 
of the central atom is now markadly asymmetric and electronic configurations 
are completely different from one another from the 16th step on. 
In the 16th elongation step, the HOMOs are still very localized on each atomic site in the 
high symmetry case, while the off-symmetry configuration shows more interatomic 
{\it d}-orbital bonding, especially along the \{111\} planes. Lastly, for strains 
larger than 33\% (steps 20 and higher), the off-axis case displays a noticeable reduction 
in the HOMO states along the thinned out part of the NW. 
For this geometry, conventional slip on \{111\} planes is not observed; instead, 
thinning occurs through a progressive ``unzipping'' process. The energy cost for a 
given strain increment is much lower for the off-symmetry case 
(Figure \ref{fig:energies}) because less energy is stored elastically.
As shown in Fig. \ref{fig:energies},
a higher engineering strain is found for the off-symmetry 
case (0.61) than for the high-symmetry one (at most 0.52), when comparing runs with 
the same stretching mechanism. The force needed to break the NW is lower for the 
slightly off-symmetry case (-1.53 nN$\pm$0.02 nN, a value close to that reported in 
experiments\cite{AgraitMD}), which is consistent with the breaking of just 
one bond instead of at least two, as occurred for the $\theta$=0 case. 

The above results suggest that abrupt (multi-bond) breaking is an unstable process and that small 
perturbations away from the symmetric case allow lower-energy atomic configurations to be reached, leading ultimately
to SAC formation.  To test this, we investigated the stability of intermediate structures
by straining initially along the [110] axis and then shifting the tensile axis to [99~101~2].
The small arrows in Fig. \ref{fig:energies} indicate the configurations 
that were used as starting points for the new $\theta$=1 simulations (whose energetics are not shown 
in the figure). Surprisingly, the symmetric deformation path is quite stable. We observed
SAC formation in only 2 cases, both of them when starting from simulations where a relatively 
high degree of atomic reorganization had occurred, and only using a relatively high 
effective strain rate. In both cases, the SAC was very short, only two atoms. All of the 
other dual-path simulations ended in abrupt breaking. These findings  
suggest that atomic configurations can be reached that are metastable.  
If perturbations are small enough, then abrupt breaking can occur. Below we will discuss this 
further, with respect to thermal fluctuations.
Lastly, we performed similar calculations
for a [111] NW elongated along the [111] tensile axis. In this case, we found SAC
formation in both the perfectly symmetric and the slightly off-symmetry cases. We
conclude that, from a thermodynamic stand point, there is an intrinsic difference in the 
evolution path of stretched gold NWs, depending on the crystallographic direction ([110] or 
[111]) of the tensile axis.

\begin{figure} 
\centering
\includegraphics[width=6.5cm]{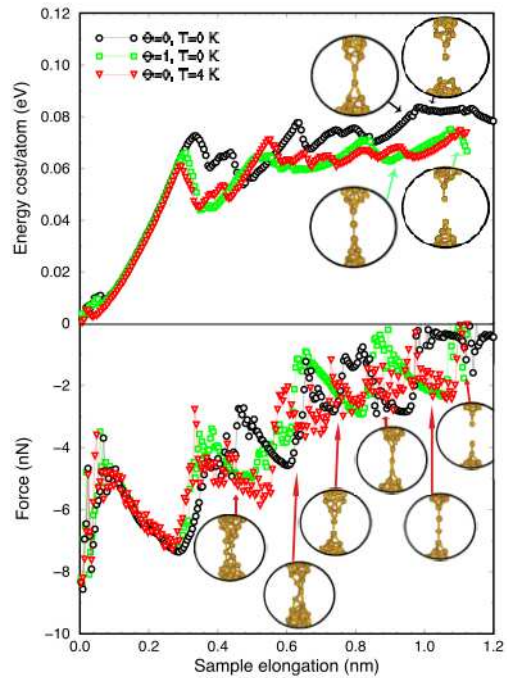}
\caption{\label{fig:Fig3} (a) Energy cost per atom versus sample elongation from
the TBMD simulations. Deformation along the [110] and [99~101~2] tensile axes is labeled ``$\theta=0$''
and ``$\theta=1$'', respectively, and the simulation temperatures are listed. 
The insets show the atomic configurations 
just before and after the NWs break, and the arrows indicate the corresponding  
elongation distance and tensile direction for T=0 K. (b) Plot of tensile forces 
as a function of sample elongation. The corresponding configurations for T=4 K 
are shown in this figure (here we show $\theta=0$, but a similar result is obtained 
for $\theta\approx 1$).  }
\end{figure}
To evaluate the impact of kinetics on the morphology and evolution of the wire, 
we performed MD simulations of the wire under stress using the TBINS 
code\cite{hri}, which is based on a
TB approximation within the
parameterization scheme proposed by Mehl and Papaconstantopoulos\cite{NRL-TB}. In this approximation, the electronic 
structure of metallic systems is fitted to the band structures and total energies obtained from
DFT calculations, as a function of volume, for face-centered, body-centered and simple-cubic 
structures. This TB scheme is also compatible with molecular dynamics.  Such parameterization has reproduced many 
static\cite{TBsurface} and dynamic\cite{TBgold,AHCG} gold properties.  
In the MD implementation of the TBINS code, 
the equations of motion of each 
atom are integrated using the Verlet-algorithm and the temperature is controlled within the Langevin scheme.
When performing our TBMD calculations, we only considered the top-and-bottom 
pulling mechanism and stretched the NW in a continuous manner, using stretching velocities of
25 m/s. These velocities are much larger than those used in typical nanocontact experiments, but still small compared 
to the speed of sound through the material, providing the system with enough time to
relax\cite{Sorensen,Bahn,AgraitMD,Ugarte2,Ugarte3,HSPark,HMS}. 
During the simulations, the atoms between the frozen slabs move and accommodate into new
configurations according to the molecular dynamics procedure.
The simulation cell is nearly identical to the one 
used above in the DFT calculations.
We considered a time step of 4 fs, and a temperature of 4 K to
simulate the experimental conditions of refs.\cite{Yanson,AgraitMD}), and 0 K 
to perform semistatic relaxation calculations akin to those
done within the DFT framework.  
During the elongation process we computed
the energy cost per atom as well as the tensile force along the stretching direction.

As before, two tensile axes are considered: parallel to the [110] and a slightly 
off-axis case (about 1 degree off the [110] direction).  Fig. 3a shows the resulting 
energy cost per atom as a function of the elongation distance of the sample.
The circle, triangle and square symbols correspond to the gold NW elongation process for 
the symmetric (T=0 K and T=4K) and the off-axis (T=0 K) cases, respectively.
The insets in this figure depict the corresponding atomic configurations just before and 
after the NW breaks for the T=0 K case, while the configurations obtained for T=4 K 
(triangle symbols) are shown in Fig. 3b.
When comparing Fig. 1 and Fig. 3a, we note that the energy costs and ultimate elongation 
distances are very similar to the DFT calculations, and that non-SAC NW breaking again 
corresponds to the higher energy cost. These results demonstrate the reliability of the TB 
approximation considered here.  Moreover, at T=0 K, a neck containing only 
two gold atoms forms for the symmetric case, while for the non-symmetric case a chain 
of three-atoms is formed. For the first case, 
both bonds below the atom in the neck stretch to accommodate the imposed strain 
and the chain breaks 
at this location. 
The T=0 K TBMD and DFT simulations therefore agree that high-symmetry stretching favors the simultaneous
breaking of multiple bonds, while
off-symmetry stretching favors SAC formation. 
On the other hand, at T=4 K, the TBMD calculations predict suspended SACs containing 
4 (see Fig. 3b) and 5 atoms (data not shown), independent of the symmetry, thus   
demonstrating that small thermal fluctuations provide enough symmetry breaking to allow 
SAC formation.

Fig. 3b shows the tensile forces as a function of sample elongation. As
observed experimentally\cite{vieira}, the tensile force increases (becomes more negative) 
with elongation and decreases when an atom is dettached from the thinning gauge section. 
Also, the force needed to detach an atom and break the NW is consistent with our DFT calculations 
and with experiments\cite{AgraitMD}.
For T=4 K, we note that the thermal fluctuations produce a noisier force curve than that 
observed for T=0 K. These fluctuations allow the system to explore a wider range 
of configurations in the potential energy surface.


In summary, we found abrupt breaking of the NWs in the completely symmetric case when thermal 
fluctuations are neglected, and a SAC when the crystal symmetry is broken 
in the early stages of the NW elongation process. 
Such crystallographic asymmetry can be induced by stretching the wire along
a slightly off-axis direction or by introducing small thermal fluctuations.
The observation of the off-axis formation of these chains 
agrees with experimental results\cite{Cadenas3}, where 
SAC formation was favored for tensile axes that were not parallel to the main crystal axis. 
We found that the distance between gold atoms within the chain ranged from 0.26 nm to 0.31 nm, 
in agreement with the bond distances reported in experiments\cite{Yanson,Cadenas3}. 
These findings represent the first quantum
calculations on the formation of suspended SACs when stretching gold
NWs along the [110] crystal orientation.
Our molecular dynamics simulations show that the static geometrical model considered
in ref.\cite{Ugarte1}
to justify the non-formation of gold monatomic chains in the [110] direction is insufficient.
Only a detailed description of the kinetics of the nanocontact elongation process can 
reproduce realistic experimental conditions.


\begin{thebibliography}{00}
\bibitem{Dis} Commercial software is identified to specify procedures.  Such identification 
does not imply recommendation by the National Institute of Standards and Technology.
\end{thebibliography}
\end{document}